\documentstyle[12pt]{article}
\begin{document}
\baselineskip = 20pt

\title{FADDEEV-JACKIW FORMALISM FOR A TOPOLOGICAL-LIKE OSCILLATOR IN
PLANAR DIMENSIONS}

\author{C. P. Natividade  \\
\it Departamento de F\'\i sica e Qu\'\i mica, Universidade Estadual
Paulista\\ 
\it Campus de Guaratinguet\'a, Caixa Postal 205\\
\it 12500-000 Guaratinguet\'a, S\~ao Paulo, BRAZIL.\\
\\ \it and\\
\\
 H. Boschi-Filho\\
\it Instituto de F\'\i sica, Universidade Federal do Rio de Janeiro \\
\it Cidade Universit\'aria, Ilha do fund\~ao, Caixa Postal 68528 \\
\it 21945-970 Rio de Janeiro, BRAZIL.\\}

\bigskip
\maketitle
\begin{abstract}
The problem of a harmonic oscillator coupling to an electromagnetic
potential plus a topological-like (Chern-Simons) massive term, in
two-dimensional space, is studied in the light of the symplectic
formalism proposed by Faddeev and Jackiw for constrained systems.
\end{abstract}
\bigskip
\bigskip
\bigskip
\vfill
\par

\pagebreak

\begin{section}{Introduction} \setcounter{equation}{0}

The dynamics of gauge field theories with the Chern-Simons
topological mass term in (2+1) dimensions \cite{DeserJT} is a quite
interesting subject and accordingly keeps being the goal of much
investigation \cite {Jackiw}. Besides its mathematical interest,
Chern-Simons theories have been an important laboratory to explain
some condensed matter phenomena like the fractional quantum Hall
effect and high-$T_c$ behavior of superconductivity \cite {DPW}. But,
perhaps, its main characteristic is that unusual spin states and
statistics appeared at the quantum mechanical level \cite{Wilczek}.
This feature has motivated the study of quantum mechanical systems in
(2+1) dimensions trying to understand the role of anyons in quantum
theory of Chern-Simons \cite{Semenoff}- \cite{Lerda}. Recently, Dunne,
Jackiw and Trugenberger \cite{DunneJT} have studied a quantum
mechanical oscillator model in close analogy to topological
Chern-Simons systems (in a reduced phase space):

\begin{equation} {L} = {B\over 2} \, \vec q \times \dot{\vec q} -
{\kappa\over 2} q^2 \;.
\end{equation}

This model basically mimics the motion of a non-relativistic point
particle in two dimensions under the influence of a perpendicular
constant magnetic field. In that paper they discussed the relation of
this model with conventional Chern-Simons theory (also with reduced
phase space so that the Maxwell term vanishes). Here, we are going to
discuss an analogous model starting from a simple charged harmonic
oscillator which experiences an external electromagnetic field
governed by a Chern-Simons term (instead of the usual Maxwell one).
The justification for such a system is that at low energies the
Chern-Simons term dominate over the Maxwell one \cite{Semenoff} and
this is the regime we are interested in (it is well known that the
inclusion of a Maxwell term besides the Chern-Simons one suppresses
fractional statistics \cite{Girotti}).

Our model can also be sought as an extension of the one discussed by
Matsuyama with canonical quantization of a charged particle in the
presence of an electromagnetic field plus a Chern-Simons term
(without an oscillator potential). A relativistic version of this
situation was also considered by Cortes, Gamboa and Velazquez
\cite{CortesGV}.  The main goal of this work is to explore the
analogy suggested in ref. \cite {DunneJT}, to study the Hamiltonian
quantization of a harmonic oscillator coupled to the electromagnetic
potential plus a topological Chern-Simons term in two dimensions,
from the symplectic formalism point of view \cite {FJ}. In this
approach, the phase space is reduced in such a way that the
Lagrangian depends on the first-order generalized velocities. The
advantage of this linearization is that the non-null Dirac brackets
are the elements of the inverse symplectic matrix \cite {HT}.
The method becomes more involved when gauge fields come together,
which is the case under analysis here, once the system gets
constrained. In this case the symplectic matrix is singular and has
no inverse unless a gauge-fixing term is included \cite {Barcelos1}.
In this work, we want to shed some light on the symplectic formalism
for constrained and unconstrained systems, taking first as an example
the simple harmonic oscillator in section 2. Section 3 is devoted to
discuss the oscillator coupling to a gauge field plus a Chern-Simons
term. We finalize the paper in section 4 with the conclusions.

\end{section}

\begin{section}{The symplectic formalism and the harmonic oscillator}
\setcounter{equation}{0}

In this section we intend to give the basic ideas of the
Faddeev-Jackiw symplectic method \cite{FJ} and discuss the
quantization of the harmonic oscillator in this scheme. For a general
review  and application to other systems see for instance refs.
\cite {Woodhouse}-\cite{Barcelos2}.

We begin considering by a phase space $\Gamma (q_i, p_i)$,
$(i=1,...,N)$ such that its algebraic structure is characterized by
the Poisson brackets

\begin{equation}
\{ q_i,q_j \} \,=\,0\,=\,\{ p_i,p_j \} \;;\;\;\;\; \{ q_i,p_j \} 
= \delta_{ij}\;.
\end{equation}

In this step, coordinates and respectively canonical momenta are
assumed to be independent variables in $\Gamma (q_i,p_i)$.

From a mathematical point of view, we can consider the coordinates of
phase space as $x^\alpha=x^\alpha (q_i,p_i)$, $(\alpha=1,...,2N)$ in
such a way that the algebraic structure is determined by a rank-two
antisymmetric tensor $\omega^{\alpha\beta}$, whose components are

\begin{equation}
\omega^{\alpha\beta}= \{ x^\alpha,x^\beta \} \;,
\end{equation}

\noindent where $\det\omega^{\alpha\beta}\not=0$.

The tensor $\omega^{\alpha\beta}$ permit us to
rewrite the Poisson bracket of two functions $A(q_i,p_i)$ and
$B(q_i,p_i)$ in a compact form

\begin{equation}
\{ A(x),B(x) \} = \partial_\alpha A
\omega^{\alpha\beta} \partial_\beta B \;,
\end{equation}

\noindent where $\partial_\alpha\equiv\partial/\partial x^\alpha$ and
since $\det\omega^{\alpha\beta}\not=0$, we can invert it. The inverse
two-form is denoted by $\omega_{\alpha\beta}$ and satisfy

\begin{equation}\label{inverse/rel}
\omega_{\alpha\beta} \omega^{\beta\gamma} = {\delta_\alpha}^\gamma\;,
\end{equation}

\noindent such that its determinant is also non-singular. A two-form
which obeys the relation (\ref{inverse/rel}) defines a symplectic
structure which, on the other hand, gives rise to generalized (Dirac)
brackets,

\begin{equation}
{\{ x^\alpha,x^\beta \}}^\ast_{GB} \equiv {\{ x^\alpha,x^\beta \}}_D 
= (\omega^{\alpha\beta})^{-1}\;.
\end{equation}

In order to show explicitly the above result, let us consider a first
order Lagrangian

\begin{equation}\label{L}
L=a_\alpha (x) {\dot x}^\alpha - V(x)\;.
\end{equation}

\noindent From the variational principle we get

\begin{equation}\label{Var.Prin}
\int dt \left[ \left( \partial_\alpha a_\beta (x) - \partial_\beta
a_\alpha (x) \right) {\dot x}^\beta - \partial_\alpha V(x) \right] =
0\; 
\end{equation}

\noindent  and
we define the two-rank antisymmetric tensor

\begin{equation}\label{matrix}
\Omega_{\alpha\beta} \equiv \partial_\alpha a_\beta  - \partial_\beta
a_\alpha \;.
\end{equation}

\noindent At this step, there are two possibilities to deal with:

{\bf a)} when $\det \Omega_{\alpha\beta} \not= 0$, we can
consider $\Omega_{\alpha\beta}$ as the symplectic matrix. In this
way, the velocities can be obtained in a trivial manner

\begin{equation}\label{velo}
{\dot x}^\alpha = \Omega^{\alpha\beta} \partial_\alpha V\;,
\end{equation}

\noindent where
$\Omega^{\alpha\beta}\equiv(\Omega_{\alpha\beta})^{-1}$. The
Hamiltonian form corresponding to eq. (\ref{velo}) is the following

\begin{equation}
{\dot x}^\alpha = \{ {x}^\alpha, V(x) \} =
{ \{ x^\alpha, x^\beta \} }_{GB} \partial_\beta V\;,
\end{equation}

\noindent therefore we can identify the generalized bracket

\begin{equation}
{\{ x^\alpha,x^\beta \} }_{GB} = \Omega^{\alpha\beta}\;.
\end{equation}

On the other hand, since there are not constraints involved in this
case (once the tensor $\Omega^{\alpha\beta}$ is invertible) we
conclude that

\begin{equation}
{\{ x^\alpha,x^\beta \} }_{GB} \equiv {\{ x^\alpha,x^\beta \} }_{D} =
\Omega^{\alpha\beta}\;,
\end{equation}

\noindent {\it i. e.}, in the symplectic formalism the Dirac brackets
are associated to the elements of the matrix $\Omega^{\alpha\beta}$
(inverse of $\Omega_{\alpha\beta}$).

Let us illustrate this unconstrained case with the example of the
(one-dimensional) harmonic oscillator

\begin{equation}
L={m\over 2} \left( {\dot q}^2 -\omega^2 q^2 \right)\;.
\end{equation}

\noindent In order to apply the symplectic formalism we should first
linearize the quadratic term ${\dot q}^2$. Following the procedure
adopted in a recent paper \cite {Barcelos2}, we have

\begin{eqnarray}\label{prescription} {\dot q}^2 \longrightarrow 2
p.{\dot q} - p^2\;.
\end{eqnarray}

\noindent Here, $p$ is an auxiliary variable. So, the first-order
Lagrangian becomes

\begin{equation}\label{L'} L=mp{\dot q} - V\;,
\end{equation}

\noindent where $V$ is the symplectic potential

\begin{equation}\label{V} V={m\over 2} \left( p^2 + \omega^2 q^2
\right)\;.
\end{equation}

\noindent Thus, from eq. (\ref{L}) we conclude that 

\begin{equation} a_q = m p\;;\;\;\;\; a_p = 0
\end{equation}

\noindent and consequently the tensor $\Omega_{\alpha\beta}$ is given
by

\begin{equation}
\Omega_{\alpha\beta}\equiv\Omega_{qp}= 
-{\partial a_q\over \partial p} + {\partial a_p\over \partial q} = -
m = - \Omega_{pq}\;.
\end{equation}

\noindent The matrix $\Omega_{\alpha\beta}$ is naturally

\begin{equation}
\Omega_{\alpha\beta}= m 
\left(
\begin{array}{cc}
0 & -1\\
+1 & 0 
\end{array}
\right)
\end{equation}

\noindent This is the symplectic matrix whose inverse permit us to
identify the brackets

\begin{equation}\label{Poisson}
\{ q,p \} ={1\over m}\;;\;\;\;\;\;  \{ q,q \} = \{ p,p \} = 0\;.
\end{equation}

The first Poisson bracket in eq. (\ref{Poisson}) has been written in
unusual form. This happened because $p$ is an auxiliary variable and
not the canonical momentum associated with variable $q$. In order to
find the usual canonical relation we should go back to the Lagrangian
(\ref{L'}). The canonical momentum is defined in the usual manner

\begin{equation}
P= {\partial L\over \partial {\dot q}} = m p\;.
\end{equation}

Therefore, we get

\begin{equation}
\{ q,P \} =1\;;\;\;\;\;\;  \{ q,q \} = \{ P,P \} = 0\;,
\end{equation}

\noindent which are the canonical Poisson bracket relations.

{\bf b)} The second possibility occurs if $\det
\Omega_{\alpha\beta}=0$. In this case we cannot identify
$\Omega_{\alpha\beta}$ as the symplectic matrix. This feature reveals
that the system under consideration is constrained \cite {Barcelos1}.
An alternative manner to circumvent this problem is to use the
constraints conveniently to change the coefficients $a_\alpha(q)$ in
the first-order Lagrangian (\ref{L}) and consequently obtain a final
two-rank tensor which could be identified with the symplectic matrix.

In the present case, we can build up an eigenvalue equation with
matrix $\Omega_{\alpha\beta}$ and $m$ $(m=1,...,M<2N)$ eigenvectors
$v^\alpha$ such that

\begin{equation}\label{zeromodes}
v^\alpha_m \Omega_{\alpha\beta} = 0\;.
\end{equation}

From eqs. (\ref{Var.Prin}) and (\ref{zeromodes}) we can write

\begin{equation}
v^\alpha_m \partial_\alpha V \equiv \Sigma_m = 0 \;,
\end{equation}

\noindent which defines possible constraints $\Sigma_m$. By imposing
that $\Sigma_m$ does not evolve in time, we arrive at

\begin{equation}
{\dot \Sigma}_m = \left(\partial_\alpha \Sigma_m \right) {\dot
q}^\alpha = 0
\end{equation}

\noindent and since ${\dot \Sigma}_m $ is linear with ${\dot
q}^\alpha$ we can incorporate this factor into the Lagrangian
(\ref{L}) by means of Lagrange multipliers $\lambda_\alpha$. So, by
considering the rescale

\begin{equation}
{\tilde a}_\alpha = a_\alpha + \lambda_\beta \partial^\beta \Sigma\;,
\end{equation}

\noindent where $a_\alpha$ is the original coefficient, we get a new
two-rank antisymmetric tensor ${\tilde \Omega}_{\alpha\beta}$ in such
a way that 

\begin{equation}
{\tilde \Omega}_{\alpha\beta} = \partial_\alpha {\tilde a}_\beta  
- \partial_\beta {\tilde a}_\alpha \;.
\end{equation}

After completing this step if $\det {\tilde \Omega}_{\alpha\beta}$ is
still vanishing we must repeat the above strategy until we find a
non-singular matrix. As has been pointed in ref. \cite{Barcelos1}, for
systems which involve gauge fields it may occur that the matrix is
singular and the eigenvectors $v^\alpha_m$ do not lead to any new
constraints. Since the main goal of this procedure is to obtain the
symplectic tensor it is necessary to choose some gauge condition. Such a
case will be discussed in next section.

\end{section}

\begin{section}{Chern-Simons Oscillator}
\setcounter{equation}{0}

Let us now extend our previous discussion to the problem of a
two-dimensional harmonic oscillator coupled to electromagnetic field
plus a Chern-Simons term. This system is described by the Lagrangian

\begin{eqnarray}
L^{(0)} & = & {m\over 2} \left[ {\dot q}_i (t) {\dot q}^i (t) 
- \omega^2 q_i (t) q^i (t) 
\right] -  \int d^2x  e A_0 (t, {\vec x}) 
\delta (\vec x - \vec q)  \nonumber\\
& + & \int d^2x e A_i (t, \vec x) \delta (\vec x - \vec q) {\dot q}^i
(t) + \theta\,\int d^2x \epsilon_{\mu\nu\rho} A^\mu (t,\vec x)
\partial^\nu A^\rho (t,\vec x)
\end{eqnarray}

\noindent where $q_i (t)$ is the particle coordinate with charge $-e$
on the plane $(i=1,2)$, $A_\mu (t,\vec x) $ is the electromagnetic
potential $(\mu=0,1,2)$,  $\theta$ is the Chern-Simons parameter,
$\epsilon_{012}=\epsilon^{012}=1$ and $g^{\mu\nu}=diag(-\,+\,+)$.
In order to proceed with the symplectic quantization of this system
we linearize the kinetic term as was done for the simple harmonic
oscillator, eq.(\ref{prescription}), so we get

\begin{equation}\label{L0}
L^{(0)} = m \left[ p_i (t) - {e\over m} A_i (t,\vec q) \right] {\dot
q}^i (t) - \theta \int \, d^2x \,  \epsilon_{ij} A^j (t,\vec x) {\dot
A}^i (t,\vec x) - V^{(0)}\;,
\end{equation}

\noindent where we used the fact that $A_i(t,\vec q)=\int d^2x
A_i (t,\vec x) \delta (\vec x -\vec q)$
and defined the potential 

\begin{equation} V^{(0)} = {m\over 2}\left[ p_i (t) p^i (t) +\omega^2
q_i (t) q^i (t) \right] + e A_0 (t,\vec q) + 2\theta \int\, d^2x\,
\epsilon_{ij}\partial^i A^j (t,\vec x) A_0 (t,\vec x)\;.
\end{equation}

Once the Lagrangian (\ref{L0}) has the general symplectic form
(\ref{L}) we can identify the coefficients

\begin{equation}\label{a0q}
a^{(0)}_{q_i} (t) = m  p_i (t) - e\, A_i (t,\vec q) 
\end{equation}

\begin{equation}\label{a0A}
a^{(0)}_{A^i} (t,\vec x) = -\, \theta\, \epsilon_{ij} A^j (t,\vec x)
\end{equation}

\begin{equation}\label{a0p}
a^{(0)}_{p_i} (t) = 0 \;;\;\;\;\;\;\;\;\;\; a^{(0)}_{A^0} = 0
\end{equation}

\noindent and calculate the matrix elements using its definition, eq.
(\ref{matrix}),

\begin{equation}\label{Omegaqp}
\Omega^{(0)}_{q_i p_j} = - m \delta_{ij}
\end{equation}

\begin{equation}\label{OmegaqA}
\Omega^{(0)}_{q_i A_j (t,\vec y)} 
= e \,\delta_{ij}\, \delta (\vec y - \vec q)
\end{equation}

\begin{equation}\label{OmegaAA}
\Omega^{(0)}_{A_i (t,\vec x) A_j (t,\vec y)} = 2\, \theta\,
\epsilon_{ij} \,
\delta (\vec x -\vec y)\;,
\end{equation}

\noindent while the others are vanishing. This way, we construct the
matrix with the convention $y^\alpha=(\vec q , \vec p , \vec A ,
A_0)$

\begin{equation}
\Omega^{(0)}_{\alpha\beta}=
\left(
\begin{array}{cccc}
0 &  -m\delta_{ij} & e \delta_{ij} \delta (\vec y - \vec q) & 0 \\
\\
m\delta_{ij} & 0 & 0 & 0 \\ 
\\
-e \delta_{ij} \delta (\vec y - \vec q) &
0 & 2 \theta \epsilon_{ij} \delta (\vec x -\vec y) & 0 \\ 
\\
0 &  0 & 0 & 0
\end{array}
\right)
\end{equation}

\medskip

\noindent which is obviously singular. Following the steps reviewed
in the previous section, we determine the non-trivial zero-modes
associated with this singular matrix, solving the equation

\begin{equation}
\Omega^{(0)}_{\alpha\beta} {v^{(0)}}^\beta = 0 \;;\;\;\;\;\;
{v^{(0)}}^\beta = 
\left(
\begin{array}{c}
a \\ 
b \\
c \\
d 
\end{array}
\right)
\;,
\end{equation}

\medskip

\noindent so that $a=b=c=0$ and $d$ remains arbitrary. Since these
zero-modes also satisfy

\begin{equation}
{v^{(0)}}^\beta \partial_\beta V^{(0)} = 0 \;,
\end{equation}

\noindent we have

\begin{equation}
d\, {\partial V^{(0)}\over \partial A_0} 
= d\, (e + 2\theta\epsilon_{ij}\partial^i A^j ) = 0
\end{equation}

\noindent and since $d$ is arbitrary we are led with a constraint:

\begin{equation}
\Sigma^{(0)} 
= e + 2\theta\epsilon_{ij}\partial^i A^j = 0\;.
\end{equation}

The next step is to remove the singularity from the symplectic matrix
by including this constraint into the Lagrangian (\ref{L0}), so we
write:

\begin{equation}\label{L1}
L^{(1)} 
= L^{(0)} + \Sigma^{(0)} \dot \lambda \;,
\end{equation}

\noindent where $\lambda=\lambda (\vec x)$ is a Lagrange multiplier
and an integration over space is assumed for the last term of the
above equation and for the following throughout.  The generalized
potential is now given by

\begin{equation}\label{V1}
V^{(1)} 
= \left. {V^{(0)}}\right|_{\Sigma^{(0)}=0} = {m\over 2} \left( p^2 +
\omega^2 q^2 \right) \;,
\end{equation}

\noindent which coincides with the one for the simple harmonic
oscillator, eq. (\ref{V}).  To calculate the new symplectic matrix we
must obtain its coefficients $a_\alpha$. In fact, they are the same
as given by eqs.  (\ref{a0q})-(\ref{a0p}) with the addition of the
one corresponding to $\lambda$:

\begin{equation}\label{a1lambda}
a_\lambda^{(1)} = e + 2\theta\epsilon_{ij}\partial^i A^j 
\end{equation}

\noindent and the exclusion of $a_{A_0}$, since $A_0$ is no longer an
explicit dynamical variable of the problem. The matrix elements are
therefore given by eqs. (\ref{Omegaqp})-(\ref{OmegaAA}) with the
addition 

\begin{equation}
\Omega^{(1)}_{A_j \lambda} =  2\theta\epsilon_{ij}\partial^i
\delta(\vec x -\vec y) 
\end{equation}

\noindent and the others are still vanishing. This way we have the
new (symplectic) matrix $(y^\alpha=(\vec q, \vec p, \vec A, \lambda))$

\begin{equation}
\Omega^{(1)}_{\alpha\beta}=
\left(
\begin{array}{cccc}
0 &  -m\delta_{ij} & e \delta_{ij} \delta (\vec y - \vec q) & 0 \\
\\
m\delta_{ij} & 0 & 0 & 0 \\ 
\\
-e \delta_{ij} \delta (\vec y - \vec q) &
0 & 2 \theta \epsilon_{ij} \delta (\vec x -\vec y) 
&  2 \theta \epsilon_{ij} \partial^i \delta (\vec x -\vec y)  \\ 
\\
0 &  0 & - 2 \theta \epsilon_{ij} \partial^j \delta (\vec x -\vec y)  
& 0
\end{array}
\right)
\end{equation}

\noindent As we can easily check this matrix is still singular.
Furthermore, the search for non-trivial zero-modes would be
unfruitful since here the potential is simply $({m\over 2})( p^2
+\omega^2 q^2)$. Therefore, following ref. \cite{Barcelos2}, we are
going to fix the gauge, which we choose to be the Weyl one $(A_0=0)$.
Once $A_0$ is absent from the Lagrangian (\ref{L1}) and noting the
equivalence $A_0=\dot\lambda$ we introduce a Lagrange multiplier
$\eta=\eta (\vec x)$ for $\lambda$:

\begin{equation}\label{L2}
L^{(2)} 
= L^{(1)} + \eta \dot \lambda \;,
\end{equation}

\noindent so that the new coefficients are given by

\begin{equation}\label{a2lambda}
a_\lambda^{(2)} = e + 2\theta\epsilon_{ij}\partial^i A^j + \eta
\;;\;\;\;\;\;\;\;\;\;\;
a_\eta^{(2)} = 0
\end{equation}

\medskip

\noindent and the others are unchanged. This way, collecting the
coefficients  we get for the symplectic matrix: $(y^\alpha=(\vec q,
\vec p, \vec A, \lambda, \eta))$

\begin{equation}
\Omega^{(2)}_{\alpha\beta}=
\left(
\begin{array}{ccccc}
0 &  -m\delta_{ij} & e \delta_{ij} \delta (\vec y - \vec q) & 0 & 0 \\
\\
m\delta_{ij} & 0 & 0 & 0 & 0 \\ 
\\
-e \delta_{ij} \delta (\vec y - \vec q) &
0 & 2 \theta \epsilon_{ij} \delta (\vec x -\vec y) 
&  2 \theta \epsilon_{ij} \partial^i \delta (\vec x -\vec y) & 0  \\ 
\\
0 &  0 & - 2 \theta \epsilon_{ij} \partial^j \delta (\vec x -\vec y)
& 0 & - \delta(\vec x -\vec y)\\
\\
0 & 0 & 0 & \delta(\vec x -\vec y) & 0
\end{array}
\right)
\end{equation}

\noindent which is not singular. Its inverse can be readily obtained
and we find

\begin{equation}
{\Omega}^{\alpha\beta} = {1\over 2m\theta}
\left(
\begin{array}{ccccc}
0 &  2\theta\delta^{ij} & 0 & 0 & 0 \\
\\
-2\theta \delta^{ij} & 0 & 
{e}, \epsilon^{ij} \delta (\vec x - \vec q) 
& 0 & 2 \theta\, e  \partial^j \delta (\vec x - \vec q)  \\ 
\\
0 & {e}  \epsilon^{ij} \delta (\vec x - \vec q) 
 & m  \epsilon^{ij} \delta (\vec x -\vec y) 
&  0 & 2m \theta  \partial^j \delta (\vec x - \vec y)  \\ 
\\
0 &  0 & 0 & 0 & 2m \theta  \delta(\vec x -\vec y)\\
\\
0 & 2\theta  {e}  \partial^i \delta (\vec x - \vec q)
 & 2m \theta  \partial^i \delta (\vec x - \vec y)
 & - 2m\theta \delta(\vec x -\vec y) & 0
\end{array}
\right)
\end{equation}

\noindent From the elements of the above matrix
we find the Dirac brackets of the theory:

\begin{equation}\label{qp}
\{ q_i,p_j \} ={1\over m} \delta_{ij}
\end{equation}

\begin{equation}\label{pA}
\{ p_i,A_j \} = {e \over 2 m \theta}\, \epsilon_{ij} 
\delta(\vec x - \vec q)
\end{equation}

\begin{equation}\label{peta}
\{ p_i, \eta \} ={ e \over m }\, \partial_{i}\, 
\delta(\vec x - \vec q)
\end{equation}

\begin{equation}\label{AA}
\{ A_i,A_j \} = { 1 \over 2 \theta}\, \epsilon_{ij} \,
\delta(\vec x - \vec y)
\end{equation}

\begin{equation}\label{Aeta}
\{ A_i, \eta \} = \partial_{i}\, \delta(\vec x - \vec y)
\end{equation}

\begin{equation}\label{lambdaeta}
\{ \lambda , \eta \} =  \,\delta(\vec x - \vec y)
\end{equation}

\noindent while the others are vanishing. The last three of the above
Dirac brackects coincide with those given by Barcelos-Neto and de
Souza \cite{Barcelos2} for the pure Chern-Simons theory, in the
$A_0=0$ gauge. The first three relations can be rewritten in terms of
the canonical momentum of the particle, as was done for the simple
oscillator in previous section:

\begin{equation}
P_i= {\partial L\over \partial {\dot q^i}} = m p_i - e A_i\;,
\end{equation}

\noindent so that we find

\begin{equation}\label{Ps}
\{ q_i,P_j \} = \delta_{ij}\;;\;\;\;
\{ p_i,A_j \} = 0\;;\;\;\;
\{ p_i, \eta \} = 0
\end{equation}

\noindent which are in agreement with Matsuyama \cite{Matsuyama}.
From the Lagrangian $L^{(2)}$, eq. (\ref{L2}), and the Euler-Lagrange
equations we find the equations of motion, which hold strongly at the
operator level:

\begin{equation}
m \dot p_i - e \dot A_i + m \omega^2 q_i = 0
\end{equation}

\begin{equation}
p_i = \dot q_i 
\end{equation}

\begin{equation}
e \dot q_i - 2 \theta \epsilon_{ji} 
(\dot A^j - \partial^j \dot \lambda) = 0
\end{equation}

\begin{equation}
\dot \eta +  2 \theta \epsilon_{ij} 
\partial^i A^j = 0
\end{equation}

\begin{equation}
\dot \lambda = 0
\end{equation}

\noindent In particular, the first two equations characterize the
harmonic oscillator motion in the presence of an external
electromagnetic field and the fourth equation defines the Lagrange
multiplier $\eta$. Substituting last equation in the third we have a
powerful relation

\begin{equation}
\dot q_i =  {2 \theta \over e}\, \epsilon_{ji} 
\, \dot A^j \;,
\end{equation}

\noindent or explicitly

\begin{eqnarray}
\dot q_i (t) & = & {2 \theta \over e}\, \epsilon_{ji} 
\,\int d^2x  \dot A^j (t,\vec x) \delta (\vec x - \vec q ) \nonumber\\
& = & {2 \theta \over e}\, \epsilon_{ji} 
\, \dot A^j (t, \vec q ) \;,
\end{eqnarray}

\noindent which can be integrated in time giving up to a constant

\begin{equation}
 A_j (t,\vec q) =  {e \over 2 \theta}\, \epsilon_{ji}  q^i (t) \;.
\end{equation}

This relation shows that the electromagnetic potential corresponds
to the one of a singular magnetic field

\begin{eqnarray}
B & = &\epsilon_{ij} \partial^i A^j  \nonumber\\
& = &\epsilon_{ij} {\partial\over \partial x_i} A^j (t,\vec q)\nonumber\\
& = & {e\over 2\theta} \delta (\vec x - \vec q)\;,
\end{eqnarray}

\noindent which gives a flux 

\begin{equation}
\phi  =  \int d^2x B = {e\over 2\theta}\;.
\end{equation}

\noindent As is well known \cite{Wilczek}, \cite{Lerda}, this
particular magnetic flux implies a fractional spin for the particle
just described, since $\theta$ can assume any value, while the
kinetic angular momentum has only integer values.

\end{section}

\begin{section}{Conclusions}

We have studied in this paper the Faddeev-Jackiw quantization for a
simple (unconstrained) oscillator and also an oscillator coupling to
a gauge field with a Chern-Simons term. This last example is
naturally constrained by virtue of the presence of gauge fields.
Apart form its academic interest we can mention that the Chern-Simons
oscillator constitutes a very interesting model, in particular for
bringing fractional statistics. Naturally, this is not surprising,
but here we have an alternative to the Dunne-Jackiw-Trugenberger
model. Our construction was also inspired in a work of Matsuyama
where a charged particle couples to electromagnetic field and
Chern-Simons term (without an oscillator potential).
The main differences from his work to ours is that he worked with
canonical quantization, in the Coulomb gauge, while we used a
symplectic formalism in another gauge ($A_0=0$). This also bring us a
bonus which indicates that in this model fractional statistics is not
a gauge artifact reaching the same conclusion as the one obtained by
Foerster and Girotti for the pure Chern-Simons theory \cite{Girotti}.
Besides, we have included a harmonic potential, which does not change
the symplectic structure so our analysis can be readily extended to
other potentials.

\vskip 2cm
\bigskip
\noindent
{\bf Acknowledgments}. One of the authors (H.B.-F.) acknowledges
interesting discussions with P. Gaete and S. Rabello. One of us
(C.P.N.) would like to acknowledge the hospitality of Universidade
Federal Fluminense were part of this work was done and FAPESP for
financial support under contract $\#$ 93/1476-3.  The authors were
partially supported by CNPq, Brazilian agency.

\end{section}

\vfill\eject


\begin{thebibliography}{99}

\bibitem{DeserJT} S. Deser, R. Jackiw and S. Templeton, Phys. Rev.
Lett. {\bf 48} (1982) 975; Annals of Phys. {\bf 140} (1982) 372.

\bibitem{Jackiw} See, for example, R. Jackiw, in ``Quantum field
theory and Quantum Statistics'', vol. 2 (1987), Ed. I.A. Batalin {\it
et al}, Bristol: Adam Hilger.

\bibitem{DPW} I. Dzyaloshinskii, A.M. Polyakov and P.B. Wiegmann,
Phys. Lett. A {\bf 127} (1988) 112; P.B. Wiegmann, Phys. Rev. Lett.
{\bf 60} (1988) 821, A.M. Polyakov, Mod. Phys. Lett. A {\bf 3} (1988)
325.

\bibitem{Wilczek} F. Wilczek, Phys. Rev. Lett. {\bf 48} (1982) 1144;
{\bf 49} (1982) 957.

\bibitem{Semenoff} G. W. Semenoff, Phys. Rev. Lett. {\bf 61} (1988) 517;
G. W. Semenoff and P. Sodano, Nucl. Phys. B {\bf 328} (1989) 753.

\bibitem{Girotti} A. Foerster and H. O. Girotti, Phys. Lett B {\bf
230} (1989) 83; Nucl. Phys. B {\bf 342} (1990) 680.

\bibitem{DunneJT} G.V. Dunne, R. Jackiw and G.A. Trugenberger, Phys.
Rev. {\bf D 41} (1990) 661.

\bibitem{Matsuyama} T. Matsuyama, J. Phys. A {\bf 23} (1990) 5241.

\bibitem{Jackiw et al} R. Jackiw and S.Y. Pi, Phys. Rev. {\bf 42}
(1990) 3500; M. Reuter, Phys. Rev. {\bf D 42} (1990) 2763; P.S. Howe
and P.K. Towsend, Class. Quant. Grav. {\bf 7} (1990) 1655; V.
Hussain, Phys. Rev. {\bf D 43} (1991) 1803.

\bibitem{CortesGV} J. L. Cortes, J. Gamboa and L. Velazquez, Phys.
Lett B {\bf 286} (1992) 105; Nucl. Phys. B {\bf 392} (1993) 645.

\bibitem{BFD} H. Boschi-Filho, C. Farina and A. de Souza Dutra, 
J. Phys. A {\bf 28} L7 (1995).

\bibitem{Lerda} For a review see A. Lerda, ``Anyons: Quantum
Mechanics of Particles with Fractional Statistics'', Springer-Verlag,
1992.

\bibitem{FJ} L. Faddeev and R. Jackiw, Phys. Rev. Lett. {\bf 60}
(1988) 1692.

\bibitem{Woodhouse} N. Woodhouse, ``Geometric Quantization'',
Claredon Press, Oxford, 1980.

\bibitem{HT} M. Henneaux and C. Teitelboin,
``Quantization  of Gauge Systems'', Princeton University Press, 1992.

\bibitem{Barcelos1} J. Barcelos-Neto and C. Wotzasek, Mod. Phys.
Lett. A {\bf 7} (1992) 1737; {\it ibid} Int. J. Mod. Phys. A {\bf 7}
(1992) 4981. 

\bibitem{Barcelos2} J. Barcelos-Neto and S.M. de Souza, Z. Phys. C
{\bf 66} (1995) 315.




\end{thebibliography}
\end{document}